\documentclass[11pt]{article}
\usepackage[utf8]{inputenc}
\usepackage{amsmath, amssymb, amsthm}
\usepackage{graphicx}
\usepackage{hyperref}
\usepackage{geometry}
\usepackage{booktabs}
\usepackage{array} 
\usepackage{parskip}
\usepackage{float}
\usepackage{booktabs}  % include in your preamble

\usepackage{float}
\usepackage{booktabs}   % prettier tables
\usepackage{multirow}   % multirow cells
\usepackage{threeparttable} % footnotes under tables
\usepackage{adjustbox}
\usepackage{xcolor}

\usepackage{booktabs}
\usepackage{adjustbox}
\usepackage{collcell} % provides \collectcell ... \endcollectcell

\newcommand{\posneg}[1]{%
  \ifdim #1 pt > 0pt
    \textcolor{green!60!black}{#1}%
  \else\ifdim #1 pt < 0pt
    \textcolor{red}{#1}%
  \else
    #1%
  \fi\fi
}

\newcolumntype{d}{>{\collectcell\posneg}c<{\endcollectcell}}

\begin{document}
\title{Evaluating the Impact of Radiographic Noise on Chest X-ray Semantic Segmentation and Disease Classification via a Scalable Noise Injection Framework}
\author{
    Derek Jiu\textsuperscript{1,2,3},
    Kiran Nijjer\textsuperscript{2,3},
    Nishant Chinta\textsuperscript{2,3},
    Ryan Bui\textsuperscript{2,3},
    Kevin Zhu\textsuperscript{3†} \\
    \\
    \textsuperscript{1}The University of Texas Health Science Center at Houston, Houston, TX, USA \\
    \textsuperscript{2}Stanford University School of Medicine, Stanford, CA, USA \\
    \textsuperscript{3}Algoverse AI Research, Palo Alto, CA, USA \\
    \texttt{derek.jiu27@gmail.com} \\
    \textsuperscript{†}Senior author
}
\maketitle

\begin{abstract}
Deep learning models are increasingly used for radiographic analysis, but their reliability is challenged by the stochastic noise inherent in clinical imaging. A systematic, cross-task understanding of how different noise types impact these models is lacking. Here, we evaluate the robustness of state-of-the-art convolutional neural networks (CNNs) to simulated quantum (Poisson) and electronic (Gaussian) noise in two key chest X-ray tasks: semantic segmentation and pulmonary disease classification. Using a scalable noise injection framework, we applied controlled, clinically-motivated noise severities to common architectures (UNet, DeepLabV3, FPN; ResNet, DenseNet, EfficientNet) on public datasets (Landmark, ChestX-ray14). Our results reveal a stark dichotomy in task robustness. Semantic segmentation models proved highly vulnerable, with lung segmentation performance collapsing under severe electronic noise (Dice Similarity Coefficient drop of 0.843), signifying a near-total model failure. In contrast, classification tasks demonstrated greater overall resilience, but this robustness was not uniform. We discovered a differential vulnerability: certain tasks, such as distinguishing \emph{Pneumothorax from Atelectasis}, failed catastrophically under quantum noise (AUROC drop of 0.355), while others were more susceptible to electronic noise. These findings demonstrate that while classification models possess a degree of inherent robustness, pixel-level segmentation tasks are far more brittle. The task- and noise-specific nature of model failure underscores the critical need for targeted validation and mitigation strategies before the safe clinical deployment of diagnostic AI.
\end{abstract}

\section{Introduction}

The proliferation of deep learning (DL) techniques has revolutionized artificial intelligence (AI) applications in medical imaging, with convolutional neural networks (CNNs) emerging as state-of-the-art approaches for classification and segmentation \cite{mienye2025deep, cai2020review, sarvamangala2021convolutional}. These architectures have shown exceptional capability in accurately identifying anatomical structures across various imaging modalities, including chest X-rays, where precise delineation and classification is critical for detecting pathological abnormalities and diagnosing life-threatening conditions such as pneumothorax, tuberculosis, and cardiovascular diseases \cite{nasser2023review, sanida2024advanced, azad2023detection}.

However, the performance of these models is heavily dependent on the quality of their training data. Despite these CNNs showcasing high accuracy when trained on curated datasets, real-world clinical X-ray imaging often contains stochastic noise—primarily quantum mottle and electronic—that degrades image quality \cite{huda2015radiographic, willemink2020preparing}, a factor many research studies overlook. Most existing research emphasizes clean-dataset optimization, thereby neglecting the distortive influence of quantum mottle and electronic noise on critical image structures \cite{willemink2020preparing, varoaux2022machine}. Emerging evidence suggests these noise sources may compromise model robustness, reducing diagnostic accuracy in AI-based segmentation and classification tasks, especially in clinics with limited resources and poor imaging quality, raising concerns about the adaptation of these models into clinical settings \cite{gebresilassie2025bridging, david-olawade2025ai-driven}. 

Although several prior studies have incorporated noise injection into deep learning models for medical imaging, most remain constrained in scope, lacking scalability, reproducibility, and systematic evaluation across diverse noise types and task domains. Past studies in this field either model Poisson–Gaussian noise for denoising/estimation without cross-task robustness tests, or assess limited noise types with ad-hoc levels and task scope \cite{lee2018poisson, jiang2023noise_robustness_ultrasound, song2024see_through_the_noise, hendrycks2019benchmarkingneuralnetworkrobustness}. 

These findings underscore a critical gap: current research lacks both a comprehensive noise injection framework and a thorough evaluation of how stochastic noise in chest X-rays affects the performance of AI-based segmentation and classification architectures. Without such validation, the clinical deployment of AI tools remains precarious in these high-stakes applications where precise segmentation is vital for diagnosis and patient care \cite{sanida2024advanced}. 

In this study, we present a scalable, noise-type–aware injection framework for systematic robustness evaluation of CNNs in chest X-ray imaging. By simulating discrete levels of Poisson (quantum) and Gaussian (electronic) noise, we generate calibrated severity gradients and measure their effects on both semantic segmentation and disease classification models. Through this framework, we quantify differential degradation trends across tasks and models, providing essential insights into the noise vulnerabilities inherent in clinical diagnostic workflows. This structured assessment lays the groundwork for more reliable AI deployment under real-world imaging variability.

\section{Key Contributions}
This work offers several key advancements in the study of noise robustness for CNNs in chest radiography:

\begin{enumerate}

  \item \textbf{Scalable intensity-parametrized noise model} Our novel noise injection framework uses a severity ladder to adjust both Poisson (quantum) and Gaussian (electronic) noise continuously. This parametric approach facilitates fine-grained robustness curves and spans scenarios from realistic low-dose to few-photon stress tests, surpassing previous fixed-level perturbation methods.
  
    \item \textbf{Empirical evaluation of differential task vulnerability} We present a comprehensive, cross-task analysis of chest X-ray segmentation and classification under calibrated Poisson and Gaussian noise severity levels. This evaluation extends beyond the single-task focus of prior studies, offering a quantitative characterization of differential vulnerabilities. Our findings reveal distinct performance degradation patterns and task-specific failure thresholds, providing an empirical foundation for understanding how segmentation and classification models respond to stochastic noise.

\end{enumerate}
These contributions address significant gaps identified in recent literature, where robust model behavior across both segmentation and classification is rarely explored, and scalable noise modeling is often absent.

\section{Related Work}

\subsection{Deep Learning and Noise in Medical Image Analysis}
Deep learning models, especially CNNs, have become the state-of-the-art for a variety of medical imaging tasks; in particular, they have proven highly effective for the classification and segmentation of anatomical structures in modalities like chest X-rays \cite{rajpurkar2017chexnetradiologistlevelpneumoniadetection, Chen_2022}. These advancements are pivotal for diagnosing numerous conditions, including tuberculosis, pneumothorax, and cardiovascular diseases. Further research has also focused on optimizing model performance through techniques like hyperparameter optimization to improve diagnostic accuracy \cite{saifullah2023detection}

\subsubsection{Noise Sources in Clinical X-ray Imaging}
Despite the power of these models, their application in real-world clinical settings is complicated by image quality degradation from stochastic noise. X-rays, albeit a fast and effective imaging modality, are prone to noise corruption, which obscures anatomical details \cite{goyal2018noise}. The two predominant forms of noise in X-ray imaging are quantum mottle and electronic noise. Quantum mottle, which is more pronounced in low-dose exposures, results from statistical fluctuations in X-ray photon detection and manifests as a grainy texture. On the other hand, electronic noise originates from thermal and electrical variations within the imaging hardware, creating a uniform haze that can obscure fine details \cite{huda2015radiographic}. The quality of chest X-ray images plays a pivotal role in automated diagnostic analysis, and factors like radiation underexposure are a frequent challenge in clinical settings \cite{rajaraman2024noise}.

\subsubsection{Impact of Noise on Model Performance}
The robustness of deep learning models to image noise is an area of growing research, with some conflicting findings. For instance, one study on pneumonia detection found no significant drop in CNN performance when Gaussian noise was added to chest radiographs \cite{kusk2023effect}. Conversely, other research highlights the sensitivity of models to data quality \cite{jang2020assessment}. It has been noted that information loss during image format conversion or the presence of ``blurry texture'' can negatively impact model performance. Labeling errors in large public datasets have also been identified as a significant issue.

\subsection{Limitations of Prior Noise Evaluation Studies}

Research addressing data corruption in medical imaging has predominantly focused on two areas: noise mitigation and, to a lesser extent, noise impact evaluation. While valuable, existing work reveals a critical gap in the systematic and scalable analysis of noise effects on deep learning model performance across diverse clinical tasks. 

Efforts in noise mitigation have yielded promising results. For instance, CNN based denoising architectures have been effectively applied to X-ray data, with some studies demonstrating superior performance when trained on experimental rather than synthetic noise \cite{chen2017cascaded}. Other innovative approaches leverage modality-specific pretext learning, pre-training models on tasks such as denoising and deblurring to bolster performance on downstream clinical applications \cite{rajaraman2024noise}. These studies underscore the importance of data quality and establish denoising as a viable mitigation strategy. 

However, effective mitigation presupposes a foundational understanding of the problem: a comprehensive characterization of how different noise sources and intensities affect model robustness is largely absent from the literature. The few studies that do inject noise to evaluate model performance often employ methodologies that are narrow in scope and lack scalability and reproducibility. For example, foundational research into multiscale filtering has been used to characterize Poisson–Gaussian noise in low-dose X-rays, but this work does not extend to evaluating the robustness of modern deep learning models across multiple tasks \cite{lee2018poisson}. 

This absence of methodological rigor is a recurring theme. While some studies have utilized synthetic quantum (Poisson) and electronic (Gaussian) noise, they often fail to implement a systematic, interval-based severity framework, typically using ad hoc severity levels that lack clinical grounding and reproducibility \cite{jiang2023noise_robustness_ultrasound, song2024see_through_the_noise}. This precludes any reproducible or comparative analysis and makes it difficult to translate findings into clinical practice. Furthermore, existing methods generally focus on a single task, such as segmentation or classification, and do not support a controlled, cross-task evaluation of noise impact.

To our knowledge, few works provide a unified, interval-defined, noise-type-aware evaluation spanning both segmentation and classification. 
We address this gap by introducing a framework that is noise-type-aware and severity-scalable, explicitly modeling quantum and electronic noise across clinically grounded, controllable severity levels. We build off this by systematically evaluating the impact of this targeted noise injection on both semantic segmentation and pulmonary disease classification, providing a comprehensive and reproducible analysis of model vulnerability in chest radiography.

\section{Methodology}

\subsection{Datasets}
Our study leverages two distinct, large-scale public datasets to ensure both anatomical and pathological diversity: a multi-institutional corpus for the segmentation task and the widely-used NIH ChestX-ray14 dataset \cite{wang2017chestxray}for classification.

\paragraph{Segmentation Dataset (Landmark)}
For semantic segmentation, we used the open-source \emph{Landmark} dataset \cite{gaggion2021landmark}, a heterogeneous collection designed to capture wide anatomical and demographic diversity. It is composed of 1,138 frontal chest radiographs aggregated from four public sources, as detailed below:

\begin{center}
\begin{tabular}{l c m{6cm}}
\hline
\textbf{Source Dataset} & \textbf{Image Count} & \textbf{Description} \\
\hline
Shenzhen Hospital (China) \cite{candemir2014lung} & 662 & Rural hospital setting; normal and pathological cases \\
PadChest (Spain) \cite{bustos2020padchest} & 200 & Large-scale; extensive labels and high variability \\
Montgomery County (USA) \cite{jang2020assessment} & 138 & Community-based radiographs for public benchmarks \\
JSRT (Japan) \cite{shiraishi2000development} & 138 & High-resolution images with detailed annotations \\
\hline
\textbf{Total} & \textbf{1,138} & \\
\hline
\end{tabular}
\end{center}

Each source provides manually annotated lung field masks created under standardized guidelines to ensure cross-dataset consistency. The full dataset was partitioned at the patient level into training, validation, and test sets with a 70/15/15 split. All images were resampled to a consistent $512 \times 512$ resolution and normalized to an intensity range of $[0,1]$ prior to any model input.

\subsubsection{Organ Stratification Tasks}
To better evaluate the generalizability of our models across distinct anatomical regions, we stratified the semantic segmentation experiments into three separate organ-specific tasks: lung-only, heart-only, and combined lung–heart segmentation. For each stratification, we retained only the relevant organ masks corresponding to the task under consideration. Datasets that provided only lung annotations, such as Shenzhen and Montgomery, were therefore included exclusively in the lung segmentation task. Datasets containing both lung and heart annotations, including JSRT and PadChest, contributed to all three stratifications. This approach allowed us to systematically compare model performance across organs of varying complexity and visibility, while ensuring that each experiment was constructed on consistent and task-appropriate ground-truth masks.

\paragraph{Classification Dataset (NIH ChestX-ray14)}
For pulmonary disease classification, we used the NIH ChestX-ray14 dataset \cite{wang2017chestxray}, which contains over 100,000 frontal chest radiographs from 30,000+ patients, labeled across 14 disease categories such as atelectasis and pneumonia. These labels are derived from automated mining of radiology reports and refined via natural language processing pipelines, as described in the original publication. 

To specifically analyze how the ambiguity of radiographic patterns influences model robustness, we curated two groups of disease-disease binary classification tasks. We define this similarity from a \textbf{computational perspective}, where pathologies are considered challenging to differentiate if they share fundamental low-level visual features (e.g., linear opacities, diffuse opacification) that can confuse an algorithmic classifier:

\begin{itemize}
  \item \textbf{Visually Similar Pairs:} Pathologies with overlapping radiographic appearances that are challenging to differentiate, such as \emph{Fibrosis vs. Pleural Thickening} and \emph{Effusion vs. Edema}. These tasks are designed to test model performance on subtle visual cues.
  \item \textbf{Visually Distinct Pairs:} Pathologies with more clearly differentiated radiographic signatures, such as \emph{Edema vs. Pleural Thickening} and \emph{Nodule vs. Effusion}. These tasks provide a baseline for model performance on less ambiguous cases.
\end{itemize}

For each curated binary task, we performed stratified splitting at the patient level to create training, validation, and test sets, ensuring no patient overlap between splits and preserving the class balance. This rigorous partitioning prevents data leakage and ensures a valid evaluation of model generalization.

\subsection{Noise Simulation and Mathematical Formulation}

\paragraph{Noise-Parameterized Input Space.}
Let $x \in \mathbb{R}^{H \times W}$ denote a normalized grayscale radiograph of height $H$ and width $W$, with pixel intensities $I(x,y) \in [0,1]$. 
We model noise-corrupted images as
\[
\tilde{x} = x + \eta(s_q, s_e),
\]
where $\eta(\cdot)$ is the resulting stochastic perturbation, which can be modeled additively for both noise types. This process is composed of two independent sources: quantum mottle noise, which results in a \textbf{signal-dependent} perturbation, and electronic readout noise, which is a \textbf{signal-independent} process. Both are parameterized by severity scalars $s_q \ge 0$ and $s_e \ge 0$, respectively. Both $s_q$ and $s_e$ are reported on a $0$--$10$ scale in our experiments, with $0$ indicating that the respective noise source is omitted. 
Larger values correspond to stronger noise perturbations.

\paragraph{Quantum noise}
Quantum noise arises from the discrete nature of photon detection and is modeled as a Poisson process. Let \(N_{0}\) denote the baseline photon count per pixel at \(s_q=1\). In our experiments, we set \(N_{0}=1000\) photons/pixel. Conventional radiographic systems typically operate on the order of \(10^{3}\)–\(10^{4}\) photons/pixel to achieve diagnostic SNR, with common low-dose imaging methods operating on the lower end of that range, at around \(10^{3}\) photons/pixel; by contrast, extreme photon-counting and few-photon studies demonstrate reconstruction feasibility at fluxes on the order of \(\sim 10\) photons/pixel \cite{zhu2018few}. 

Hence, we parameterize a dose/severity scalar \(s_q\ge 1\) so that the effective photons per pixel decrease with increasing severity according to
\[
N_{\mathrm{ph}}(s_q) \;=\; \frac{N_{0}}{s_q^{2}}.
\]
This choice is motivated by basic Poisson statistics: if a pixel receives on average \(N\) photons, the Poisson variance equals the mean and the photon-noise SNR scales as
\[
\mathrm{SNR}_{\text{photon}} \;=\; \frac{N}{\sqrt{N}} \;=\; \sqrt{N},
\]
i.e. SNR \(\propto \sqrt{N}\). Thus, to obtain a linear inverse dependence of SNR on the severity parameter \(s_q\) (so that doubling \(s_q\) halves the SNR), we set \(N \propto 1/s_q^{2}\). The algebraic consequence of this mapping is straightforward and is shown below \cite{zhu2018few}.

For a normalized image intensity \(I(i,j)\in[0,1]\) (a transmission proxy), the Poisson parameter at pixel \((i,j)\) is
\[
\lambda(i,j; s_q) \;=\; I(i,j)\,N_{\mathrm{ph}}(s_q)
= I(i,j)\,\frac{N_0}{s_q^{2}}.
\]
We draw
\[
P(i,j) \sim \mathrm{Poisson}\big(\lambda(i,j; s_q)\big),
\]
and rescale to an intensity estimate
\[
Q(i,j) \;=\; \frac{P(i,j)}{N_{\mathrm{ph}}(s_q)}.
\]
By standard Poisson properties \(\mathbb{E}[P]=\lambda\) and \(\mathrm{Var}[P]=\lambda\), so
\[
\mathbb{E}\big[Q(i,j)\big] = I(i,j),
\qquad
\mathrm{Var}\big[Q(i,j)\big] = \frac{\lambda(i,j; s_q)}{N_{\mathrm{ph}}(s_q)^{2}}
= I(i,j)\,\frac{s_q^{2}}{N_0}.
\]
Hence the quantum-only SNR (per pixel, before any other noise sources) is
\[
\mathrm{SNR}_{\mathrm{quantum}}(i,j)
\;=\; \frac{\mathbb{E}[Q(i,j)]}{\sqrt{\mathrm{Var}[Q(i,j)]}}
\;=\; \frac{I(i,j)}{\sqrt{I(i,j)\,s_q^{2}/N_0}}
\;=\; \frac{\sqrt{I(i,j)\,N_0}}{s_q}.
\]

Because \(\mathrm{SNR}_{\mathrm{quantum}}\propto 1/s_q\) under this parametrization, \(s_q\) acts as an intuitive severity scalar: increasing \(s_q\) reduces photon flux and therefore reduces SNR in a controlled, interpretable way. The above Poisson–Gaussian modeling framework and parameter-estimation approaches are standard in low-dose X-ray imaging literature \cite{lee2018poisson}.

\paragraph{Interpretation and experimental ladder}
With \(N_0=1000\) photons/pixel at \(s_q=1\), the mapping yields the following example photon budgets:

\begin{center}
\begin{tabular}{c c c}
\hline
$s_q$ & $N_{\mathrm{ph}}(s_q)$ (photons/pixel) & regime \\
\hline
1 & 1000 & realistic low-dose \\
2 & 250 & low (noticeable shot noise) \\
4 & 62.5 & moderate shot noise \\
6 & 27.8 & strong shot noise \\
8 & 15.6 & very low (extreme degradation) \\
10 & 10 & few-photon (extreme stress test) \\
\hline
\end{tabular}
\end{center}

We emphasize that the highest-severity settings (e.g., \(s_q\gtrsim 10\)) enter the few-photon regime explored in photon-counting research and are therefore best interpreted as controlled stress tests for model robustness rather than direct models of routine clinical exposure \cite{zhu2018few}.

\paragraph{Electronic noise}
We model electronic readout noise as additive, signal-independent Gaussian perturbations, reflecting its origin from thermal fluctuations in the detector electronics \cite{samei1998practical}. While the global magnitude of this noise is typically low in well-calibrated systems, its diagnostic impact becomes dominant in low-signal regions of an image, such as behind the mediastinum or spine, and in low-resource environments with outdated equipment or limited maintenance capabilities. Our framework is designed to simulate the visual appearance of these challenging, electronic-noise-dominated conditions.

Concretely, let $\sigma_{0}$ denote a baseline standard deviation (we fix $\sigma_{0}=0.1$ in normalized intensity units).
An electronic intensity parameter $s_e \ge 0$ scales the noise level:
\[
\sigma_e(s_e) = \sigma_0 \, s_e, \qquad \varepsilon(x,y) \sim \mathcal{N}\big(0, \sigma_e(s_e)^2\big).
\]
The corrupted image is
\[
E(x,y) = I(x,y) + \varepsilon(x,y),
\]
with variance contribution $(\sigma_0 s_e)^2$.

\paragraph{Interpretation and experimental ladder} With $\sigma_{0}=0.1$, the severity ladder is linear in:
\[
s_{e}\in\{0,1,2,4,6,8,10\}\quad\Rightarrow\quad \sigma_{e}\in\{0,0.1,0.2,0.4,0.6,0.8,1.0\}\ \text{(normalized units)}.
\]

The lower end of our ladder ($s_e \le 2$) can be interpreted as simulating the effect of a globally noisy detector due to factors like high temperature or electromagnetic interference. The higher end ($s_e \ge 8$) provides a clinically-grounded stress test, simulating the severe, low signal-to-noise ratio conditions encountered when analyzing faint structures in the most heavily attenuated, electronic-noise-dominated regions of a chest radiograph \cite{samei1998practical}. These high values induce visible saturation, intentionally pushing models to their failure point to comprehensively map their robustness boundaries.

\begin{figure}[h] % 'h' means place "here" if possible
    \centering
    \includegraphics[width=0.8\textwidth]{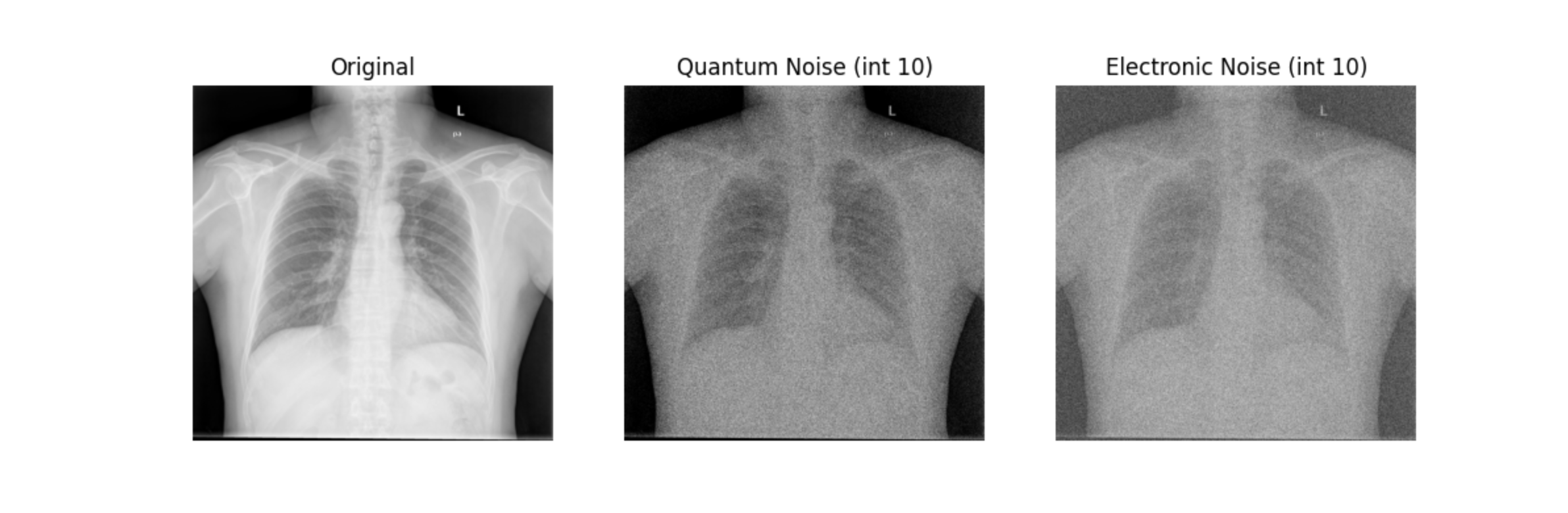} % Without file extension
    \caption{An original chest radiograph (left) is shown with the addition of maximum-intensity quantum noise ($s_q$ = 10, middle) and electronic noise ($s_e$ = 10, right) to demonstrate their distinct visual textures.  Quantum noise produces a blotchy, signal-dependent mottle, while electronic noise introduces a fine-grained, uniform static.}
    \label{fig:example}
\end{figure}

By standard Gaussian properties, $\mathbb{E}[E(x,y)] = I(x,y)$ and 
$\mathrm{Var}[E(x,y)] = \sigma_e(s_e)^2 = (\sigma_0 s_e)^2$. 
Hence the electronic-only SNR (per pixel, before any other noise sources) is
\[
\mathrm{SNR}_{\mathrm{electronic}}(i,j)
\;=\; \frac{\mathbb{E}[E(i,j)]}{\sqrt{\mathrm{Var}[E(i,j)]}}
\;=\; \frac{I(i,j)}{\sigma_0 s_e}.
\]

While both SNRs are inversely proportional to their respective severity parameters ($1/s_q$ and $1/s_e$), their dependence on the image signal $I(i,j)$ is fundamentally different. Quantum noise is most pronounced in bright, high-signal regions (where its variance scales with $\sqrt{I}$), whereas the impact of electronic noise is most significant in faint, low-signal structures where the signal $I$ is small relative to the fixed noise floor $\sigma_e$.

In both noise source implementations, the image is normalized to [0,1] before noise injection and re-converted to 8-bit format afterward. This formulation enables separate manipulation of quantum and electronic noise but is calibrated via heuristic constants rather than physical measurements.

\paragraph{Segmentation task}
For semantic segmentation, each model implements a mapping
\[
f_{\theta}: \mathbb{R}^{H \times W} \to [0,1]^{H \times W \times C},
\]
where $C$ is the number of anatomical classes (including background), and $f_{\theta}(\tilde{x})_{i,j,c}$ denotes the predicted probability that pixel $(i,j)$ belongs to class $c$.  
We train $\theta$ by minimizing the soft Dice loss $\mathcal{L}_{\mathrm{seg}}$:  
\[
\theta^{*} = \arg\min_{\theta} \ \mathbb{E}_{(x, y) \sim \mathcal{D}} \ \mathcal{L}_{\mathrm{seg}}\big( f_{\theta}(\tilde{x}), y \big),
\]

\paragraph{Classification task}
For disease classification, each model defines a mapping
\[
g_{\phi}: \mathbb{R}^{H \times W} \to \Delta^{K-1},
\]
where $\Delta^{K-1}$ is the probability simplex over $K$ disease classes.  
The parameters $\phi$ are optimized with categorical cross-entropy $\mathcal{L}_{\mathrm{cls}}$ :
\[
\phi^{*} = \arg\min_{\phi} \ \mathbb{E}_{(x, y) \sim \mathcal{D}} \ \mathcal{L}_{\mathrm{cls}}\big( g_{\phi}(\tilde{x}), y \big).
\]

\subsection{Model Architectures and Pre-training}
To evaluate the impact of stochastic noise, we selected a suite of state-of-the-art CNN architectures representing different design paradigms for both semantic segmentation and classification. A cornerstone of our methodology is the use of transfer learning; all models were initialized with weights pre-trained on the ImageNet dataset to establish powerful and clinically-relevant performance baselines.

\paragraph{Segmentation Models}
For the segmentation task, we employed encoder-decoder models from the \texttt{segmentation-models-pytorch} library. To create a controlled comparison, all segmentation models used the same feature extractor: a \textbf{ResNet34} encoder \cite{he2016deep}. This allows us to isolate the impact of the decoder's architectural design—the component responsible for upsampling features and generating the final pixel-level mask—on noise robustness. We evaluated three canonical decoder architectures:
\begin{itemize}
    \item \textbf{UNet:} Employs a symmetric encoder-decoder structure with skip connections that pass fine-grained spatial information from the encoder to the decoder, renowned for its precise localization capabilities \cite{ronneberger2015u}.
    \item \textbf{DeepLabV3:} Utilizes atrous spatial pyramid pooling (ASPP) to probe features at multiple scales with parallel atrous convolutions, allowing the model to capture rich multi-scale context \cite{chen2017rethinkingatrousconvolutionsemantic}.
    \item \textbf{Feature Pyramid Network (FPN):} Constructs a rich, multi-scale feature pyramid from the encoder's feature hierarchy, combining high- and low-level features to improve the detection of objects at different scales \cite{lin2017feature}.
\end{itemize}

\paragraph{Classification Models}
For the classification task, we evaluated the backbone architectures directly: \texttt{ResNet34}, \texttt{DenseNet} \cite{huang2017densely}, and \texttt{EfficientNet} \cite{tan2019efficientnet}. This provides a parallel analysis of how these feature extractors perform on a global, image-level task compared to their role within a segmentation model.

\subsection{Experimental Protocol and Training}
Our protocol was designed to rigorously evaluate the inherent robustness of these pre-trained models. The methodology involves fine-tuning each model on our clean, noise-free training data, followed by a systematic evaluation using our noise injection framework during inference.

\paragraph{Fine-tuning and Optimization}
Consistent with best practices for transfer learning, all models were fine-tuned using the AdamW optimizer with an initial learning rate of $1\times 10^{-4}$. The learning rate was decayed over a maximum of 200 epochs via a cosine annealing schedule. To ensure training stability, we employed gradient clipping with a maximum norm of 0.5. Furthermore, we utilized Stochastic Weight Averaging (SWA) to improve model generalization by averaging the model's weights over the final training stages, which often finds a wider and more robust solution minimum. To prevent overfitting and ensure the selection of the best performing model, we employed an early stopping strategy, halting training if the validation loss did not improve for a patience of 10 consecutive epochs. The model checkpoint corresponding to the best performance on the clean validation set was saved for the final evaluation. 

\paragraph{Data Augmentation}
Standard data augmentation techniques were applied during fine-tuning to improve model generalization. For classification, the training set was augmented using random horizontal flips and mild affine transformations (rotation $\pm 10^\circ$). For segmentation, we used random horizontal flips to create a robust baseline without corrupting fine anatomical boundary details.

\paragraph{Implementation Details}
Experiments were conducted in PyTorch and PyTorch Lightning. To match the 3-channel input requirement of the ImageNet pre-trained encoders, single-channel grayscale X-ray images were triplicated. A binary cross-entropy with logits loss was used for classification, and a soft Dice loss for segmentation. We assume the two noise sources are statistically independent. Our complete code will be made publicly available to ensure full reproducibility.

\section{Results and Discussion}

In both the classification and segmentation tasks, performance trends were highly consistent across the evaluated architectures (ResNet, DenseNet, EfficientNet and UNet, DeepLabV3, FPN, respectively). This consistency suggests that the observed effects are primarily attributable to the fundamental challenges posed by the noise, rather than idiosyncratic model behaviors. Consequently, to isolate the impact of noise modality and intensity, all results are reported as the mean across the three architectures for each task.

\subsection{Segmentation Results}

In the semantic segmentation tasks, we observed a clear hierarchy of vulnerability, where the anatomical target's characteristics—such as size and boundary definition—dictated its resilience to stochastic noise.

\paragraph{Lung Segmentation}
The task of segmenting the lungs, a large structure with generally well-defined boundaries, was moderately impacted by quantum noise. At maximum quantum severity ($s_q=10$), the mean Dice Similarity Coefficient (DSC) and Intersection-over-Union (IoU) decreased by 0.387 and 0.396, respectively. However, this task proved critically vulnerable to electronic noise. At maximum electronic severity ($s_e=10$), performance collapsed, with DSC and IoU plummeting by 0.843 and 0.875, respectively, signifying a near-total failure of the models to identify the lung fields.

\begin{table}[htbp]
\centering
\begin{threeparttable}
\begin{adjustbox}{max width=\textwidth}
\begin{tabular}{cc|c d c d | c d c d | c d c d | c d c d}
\toprule
 & &
 \multicolumn{4}{c|}{Landmark (Heart)} &
 \multicolumn{4}{c|}{Landmark (Lungs \& Heart)} &
 \multicolumn{4}{c|}{Landmark (Lungs)} &
 \multicolumn{4}{c}{Shenzhen (Lungs)} \\
\cmidrule(lr){3-6} \cmidrule(lr){7-10} \cmidrule(lr){11-14} \cmidrule(lr){15-18}
\multicolumn{1}{c}{\shortstack{Quantum\\Intensity}} &
\multicolumn{1}{c}{\shortstack{Electronic\\Intensity}} &
\multicolumn{1}{c}{Dice} &
\multicolumn{1}{c}{\shortstack{Dice\\Diff\tnote{1}}} &
\multicolumn{1}{c}{IOU} &
\multicolumn{1}{c|}{\shortstack{IOU\\Diff\tnote{1}}} &

\multicolumn{1}{c}{Dice} &
\multicolumn{1}{c}{\shortstack{Dice\\Diff\tnote{1}}} &
\multicolumn{1}{c}{IOU} &
\multicolumn{1}{c|}{\shortstack{IOU\\Diff\tnote{1}}} &

\multicolumn{1}{c}{Dice} &
\multicolumn{1}{c}{\shortstack{Dice\\Diff\tnote{1}}} &
\multicolumn{1}{c}{IOU} &
\multicolumn{1}{c|}{\shortstack{IOU\\Diff\tnote{1}}} &

\multicolumn{1}{c}{Dice} &
\multicolumn{1}{c}{\shortstack{Dice\\Diff\tnote{1}}} &
\multicolumn{1}{c}{IOU} &
\multicolumn{1}{c}{\shortstack{IOU\\Diff\tnote{1}}} \\
\midrule
0.00 & 0.00 & 0.86 & 0.00 & 0.76 & 0.00 & 0.89 & 0.00 & 0.81 & 0.00 & 0.86 & 0.00 & 0.76 & 0.00 & 0.88 & 0.00 & 0.79 & 0.00 \\
0.00 & 1.00 & 0.84 & -0.02 & 0.73 & -0.03 & 0.83 & -0.06 & 0.71 & -0.10 & 0.84 & -0.02 & 0.73 & -0.03 & 0.90 & 0.02 & 0.82 & 0.03 \\
0.00 & 2.00 & 0.93 & 0.07 & 0.88 & 0.12 & 0.83 & -0.06 & 0.72 & -0.09 & 0.82 & -0.04 & 0.70 & -0.06 & 0.90 & 0.02 & 0.82 & 0.03 \\
0.00 & 4.00 & 0.72 & -0.14 & 0.57 & -0.19 & 0.69 & -0.21 & 0.52 & -0.28 & 0.72 & -0.14 & 0.57 & -0.19 & 0.93 & 0.05 & 0.87 & 0.08 \\
0.00 & 6.00 & 0.64 & -0.22 & 0.48 & -0.28 & 0.73 & -0.16 & 0.58 & -0.23 & 0.63 & -0.23 & 0.47 & -0.29 & 0.86 & -0.02 & 0.76 & -0.03 \\
0.00 & 8.00 & 0.79 & -0.08 & 0.65 & -0.11 & 0.78 & -0.08 & 0.65 & -0.11 & 0.78 & -0.02 & 0.91 & -0.03 & 0.77 & -0.12 & 0.62 & -0.17 \\
0.00 & 10.00 & 0.59 & -0.27 & 0.48 & -0.28 & 0.23 & -0.63 & 0.13 & -0.63 & 0.13 & -0.84 & 0.07 & -0.88 & 0.60 & -0.28 & 0.43 & -0.36 \\
1.00 & 0.00 & 0.92 & 0.06 & 0.85 & 0.09 & 0.84 & -0.05 & 0.73 & -0.08 & 0.90 & -0.07 & 0.82 & -0.12 & 0.93 & 0.04 & 0.86 & 0.07 \\
1.00 & 1.00 & 0.71 & -0.16 & 0.54 & -0.22 & 0.83 & -0.06 & 0.71 & -0.10 & 0.86 & 0.00 & 0.76 & 0.00 & 0.85 & -0.03 & 0.74 & -0.05 \\
2.00 & 0.00 & 0.89 & 0.02 & 0.80 & 0.04 & 0.83 & -0.06 & 0.71 & -0.10 & 0.84 & -0.02 & 0.71 & -0.05 & 0.93 & 0.05 & 0.88 & 0.09 \\
4.00 & 0.00 & 0.93 & 0.07 & 0.87 & 0.11 & 0.81 & -0.08 & 0.68 & -0.13 & 0.72 & -0.14 & 0.59 & -0.17 & 0.93 & 0.05 & 0.87 & 0.08 \\
6.00 & 0.00 & 0.86 & -0.01 & 0.75 & -0.01 & 0.63 & -0.26 & 0.46 & -0.35 & 0.54 & -0.32 & 0.55 & -0.21 & 0.92 & 0.04 & 0.86 & 0.07 \\
8.00 & 0.00 & 0.88 & 0.01 & 0.78 & 0.02 & 0.85 & -0.04 & 0.74 & -0.06 & 0.50 & -0.36 & 0.43 & -0.33 & 0.88 & -0.01 & 0.78 & -0.01 \\
10.00 & 0.00 & 0.90 & 0.04 & 0.82 & 0.06 & 0.78 & -0.11 & 0.64 & -0.17 & 0.47 & -0.39 & 0.36 & -0.40 & 0.66 & -0.23 & 0.49 & -0.30 \\
\bottomrule
\end{tabular}
\end{adjustbox}
\begin{tablenotes}
\footnotesize

\item[1] Difference from the baseline (Quantum Intensity = 0, Electronic Intensity = 0), shown in Row 1.
\end{tablenotes}
\end{threeparttable}
\caption{Average semantic segmentation performance (DSC and IoU) across varying quantum and electronic noise severities. The first three columns report results on the full Landmark dataset (covering lungs only, combined lung-heart, and heart only tasks). The final columns show performance on the Shenzhen dataset alone to demonstrate cross-dataset generalization.}
\end{table}

\paragraph{Heart Segmentation}
In stark contrast, segmentation of the heart was remarkably resilient, particularly to quantum noise. Counter-intuitively, at maximum quantum severity, DSC and IoU showed a marginal \emph{increase} of 0.04 and 0.06. This suggests that the high-frequency nature of quantum noise does not disrupt the salient features required for cardiac boundary detection. This resilience did not fully extend to electronic noise, which induced a moderate performance decline, with DSC and IoU reducing by 0.269 and 0.278, respectively.

\paragraph{Lung-Heart Segmentation}
When segmenting both structures simultaneously, performance degradation fell between the two single-organ scenarios. Maximal quantum noise resulted in DSC and IoU decreases of 0.112 and 0.176—more severe than heart-only but far less than lung-only. Consistent with the other tasks, electronic noise was far more detrimental, reducing DSC and IoU by 0.631 and 0.634, highlighting the compounded challenge of multi-structure segmentation in the presence of severe electronic noise.

\paragraph{Anomalous Performance Recovery}
A noteworthy trend emerged across all segmentation tasks in response to electronic noise. While the overarching pattern was degradation with increasing severity, a slight but consistent performance recovery was observed at an intensity of $s_e=8$ relative to $s_e=6$, before resuming its sharp decline at $s_e=10$. We hypothesize this may be a form of stochastic regularization, where intermediate noise disrupts the model's reliance on spurious textural artifacts, temporarily improving generalization before signal corruption becomes overwhelming. This non-monotonic behavior suggests a complex interplay between noise-as-regularizer and noise-as-corruptor.

\subsection{Classification Results}

In contrast to the large performance collapse observed in the lung segmentation task, our classification models demonstrated a markedly greater degree of overall robustness. While still susceptible to significant, noise-induced performance degradation, the models' predictive capabilities generally degraded more gracefully and did not suffer the same near-total failure, revealing a more nuanced and task-dependent relationship with noise.

\begin{table}[htb]
  \centering
  \label{tab:baseline}
  \begin{tabular}{l c c c}
    \toprule
    Disease Pair & AUPRC & AUROC & F1 Score \\
    \midrule
    \multicolumn{4}{l}{\textbf{Visually Similar}} \\
    Fibrosis and Pleural Thickening & 0.680 & 0.550 & 0.753 \\
    Effusion and Edema              & 0.323 & 0.740 & 0.379 \\
    Pneumothorax and Atelectasis    & 0.828 & 0.737 & 0.810 \\
    \midrule
    \multicolumn{4}{l}{\textbf{Visually Distinct}} \\
    Fibrosis and Infiltration       & 0.948 & 0.605 & 0.963 \\
    Edema and Pleural Thickening    & 0.919 & 0.881 & 0.880 \\
    Nodule and Effusion             & 0.686 & 0.706 & 0.694 \\
    \bottomrule
  \end{tabular}
  \caption{Baseline average performance of the fine-tuned classification models across the six visually similar and visually distinct disease pairs, under noise‑free conditions. Performance metrics vary significantly, indicating different intrinsic task difficulties.}
  \label{tab:mylabel}
\end{table}

\begin{figure}[H]
    \centering
    \includegraphics[width=0.95\textwidth]{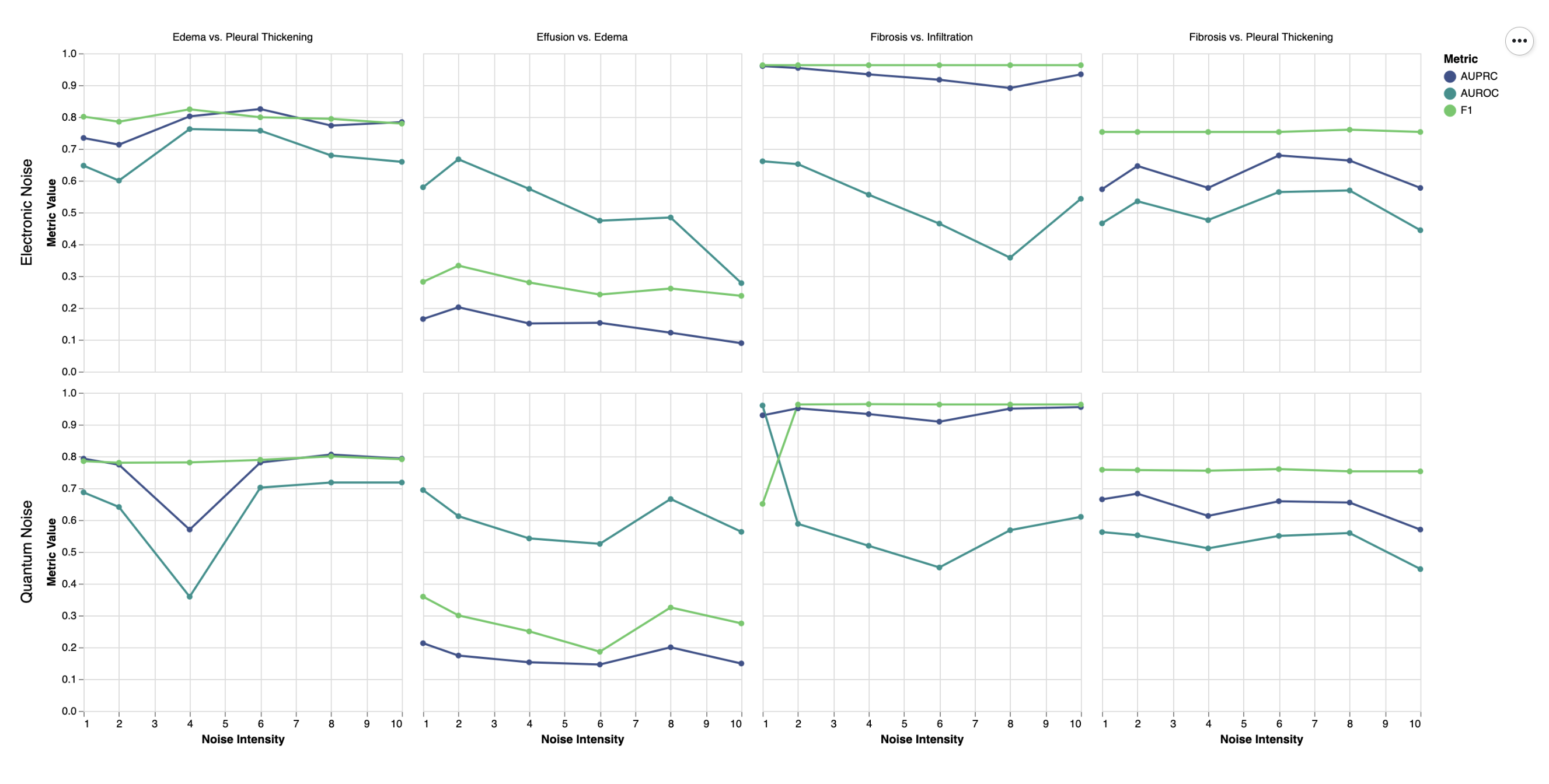} % Without file extension
    \caption{Average classification performance (AUPRC, AUROC, F1) is plotted against quantum ($s_q$) and electronic ($s_e$) noise severity. Two representative disease pairs from each category (visually similar and distinct) are shown to illustrate the differential impact of noise, where model degradation is highly dependent on both the clinical task and the noise modality. While all three metrics are displayed for completeness, the discussion in the main text focuses on AUROC as the primary metric.}
    \label{fig:graphs}
\end{figure}

\paragraph{Visually Similar Pairs}
Under noise-free conditions, the models showed vastly different capabilities across the three visually similar disease pairs, highlighting their varying intrinsic difficulties (Figure \ref{tab:baseline}). The model struggled significantly to differentiate \emph{Fibrosis from Pleural Thickening}, achieving an AUROC of just 0.550. In contrast, it performed competently on \emph{Effusion vs. Edema} (AUROC 0.740) and was strongest on \emph{Pneumothorax vs. Atelectasis} (AUROC 0.737, AUPRC 0.828). This wide disparity provides crucial context for interpreting the subsequent effects of noise. The central finding of our classification experiment is that models were moderately susceptible to noise but exhibited a relative resilience not seen in segmentation. This overall robustness, however, was punctuated by cases of \emph{differential vulnerability}, where specific tasks were uniquely sensitive to different types of noise (Figure \ref{fig:graphs}). This resilience was not universal. The \emph{Pneumothorax vs. Atelectasis} task, for example, proved exceptionally sensitive to quantum noise. At maximum intensity ($s_q=10$), its AUROC plummeted by 0.355 to 0.382, performing worse than random guessing. However, the same task was remarkably robust to electronic noise, with its AUROC declining by a more modest 0.123. This opposing pattern was also observed for the \emph{Effusion vs. Edema} task, which was highly vulnerable to electronic noise (AUROC drop of 0.462) but less affected by quantum noise.

\paragraph{Visually Distinct Pairs}
The trend of relative robustness was even more evident on disease pairs with more distinct radiographic features. As hypothesized, baseline performance was generally higher (Figure \ref{tab:baseline}), with the \emph{Edema vs. Pleural Thickening} task achieving a strong AUROC of 0.881. 

Despite this stronger starting point, models remained moderately susceptible to noise. At maximal quantum severity ($s_q=10$), the average AUPRC across these tasks dropped by 0.178 and the AUROC by 0.212. Electronic noise produced a similar impact, with average AUPRC and AUROC declining by 0.138 and 0.230, respectively, at $s_e=10$. 

Crucially, however, even at these maximum noise severities, the average AUROC across these tasks remained well above the random-chance threshold. This stands in stark contrast to the more pronounced performance collapse observed in the organ-stratified segmentation tasks under similar conditions. This finding suggests that global, image-level classification tasks possess an inherent architectural or feature-level robustness that is absent in pixel-level segmentation tasks.

\section{Limitations and Future Work}
While our framework enables controlled and reproducible robustness testing, this study has several limitations that present valuable opportunities for future research.

\begin{enumerate}

  \item \textbf{Fidelity of the Noise Simulation:} Our noise synthesis operates on post-processed, normalized images rather than raw detector data. While calibrated to span realistic clinical scenarios, this limits the direct physical interpretability of our severity levels in terms of radiation dose or absolute photon counts. Furthermore, we test quantum and electronic noise independently and use simplified distributions, whereas real-world images contain a complex mixture of simultaneously present noise sources, including scatter, fixed-pattern noise, and other artifacts. 
  
  \emph{Future direction:} A crucial next step is to validate our synthetic severity ladder against real low-dose clinical or phantom data. Future iterations of the framework could also implement a more complex mixed-noise model (e.g., Poisson-Gaussian) and incorporate spatially correlated or non-ideal noise sources to achieve even higher physical fidelity.

  \item \textbf{Architectural Scope:} Our study focused on a representative set of canonical CNN-based decoders, all built upon a ResNet34 backbone. While this controlled setup was essential for isolating the impact of decoder design, the findings may not generalize to all architectural paradigms. The recent rise of Vision Transformers (ViTs) and other non-convolutional architectures presents a new frontier for robustness analysis. 
  
  \emph{Future direction:} Future work should extend this benchmark to include a wider range of backbones (e.g., larger ResNets, EfficientNets) and entirely different architectural families, particularly ViTs, to investigate whether their global attention mechanisms offer inherent advantages in robustness to stochastic noise.

  \item \textbf{From Evaluation to Mitigation:} This work focuses on quantifying the problem of noise-induced degradation. The logical next step is to investigate solutions. We did not explore mitigation strategies that could potentially harden these models against the vulnerabilities we identified. 
  
  \emph{Future direction:} Our framework provides the ideal testbed for evaluating various defense mechanisms. Future research should assess the efficacy of noise-robust training techniques (e.g., adversarial training, targeted data augmentation with our noise model) or the use of deep learning-based denoising models as a pre-processing step to improve downstream task performance.

  \item \textbf{Explaining Differential Vulnerability:} One of our most significant findings was the differential vulnerability of specific tasks to different noise types. For instance, the Pneumothorax vs. Atelectasis task was highly sensitive to quantum noise, while the Effusion vs. Edema task was more vulnerable to electronic noise. Our current work hypothesizes the reasons for this but does not provide a mechanistic explanation. 
  
  \emph{Future direction:} Investigating the underlying cause of this phenomenon is a rich area for future work. Techniques from explainable AI (XAI), such as saliency mapping (e.g., Grad-CAM) under different noise conditions, could be used to determine if models rely on high-frequency edge features (vulnerable to quantum noise) for one task and low-frequency textural features (vulnerable to electronic noise) for another.

\end{enumerate}

Systematically addressing these limitations is a critical step toward the safe and effective translation of these models into routine clinical practice.

\section{Conclusion}

We present a systematic, cross-task evaluation of how simulated quantum and electronic noise impacts CNNs for chest X-ray semantic segmentation and pulmonary disease classification. Our work reveals that the impact of noise is far from monolithic, with global, image-level classification tasks demonstrating significantly greater resilience compared to the more brittle failure modes of pixel-level segmentation. 

Our analysis uncovered a key finding of \emph{differential vulnerability}: certain diagnostic tasks were catastrophically sensitive to high-frequency quantum noise, while others were more susceptible to the broader artifacts of electronic noise. This task-dependency was mirrored in segmentation, where the large, well-defined lung fields proved highly vulnerable to both noise types, while the anatomically distinct heart region was remarkably robust. Furthermore, we observed a consistent, non-monotonic degradation pattern across tasks, suggesting a complex interplay where intermediate noise may act as a regularizer before becoming overwhelmingly corruptive. 

These nuanced findings were made possible by our novel parameterized severity ladder, which provides a reproducible tool for stress-testing models beyond simple data augmentation. Our framework spans from realistic low-dose imaging conditions to the upper bound of clinically observed noise levels, enabling the robust benchmarking needed to bridge the critical gap between idealized training environments and the stochastic realities of real-world clinical deployment. By providing a more compact and actionable picture of noise-driven failure modes, our work paves the way for the development of more reliable and robust diagnostic AI models.

\section*{Conflicts of Interest}
The authors declare that they have no competing interests.

\section*{Data and Code Availability}
The code and noise-injection framework will be made publicly available via GitHub upon publication. All datasets used are publicly accessible (Landmark composite and NIH ChestX-ray14).

\clearpage
\bibliographystyle{unsrt}
\bibliography{references}
\end{document}